\documentclass{svproc}
\usepackage{url}

\usepackage{bm}       
\usepackage{amsmath}
\usepackage{amssymb}
\usepackage{graphicx}

\newcommand{\ket}[1]{ | #1 \rangle}

\begin{document}
\mainmatter              
\title{Pion-cloud contribution to the $N\rightarrow \Delta$ transition form factors}
\titlerunning{Cluster Separability}  
%
\author{Ju-Hyun~Jung \and Wolfgang~Schweiger}
\authorrunning{J.-H. Jung} 
%
%
\institute{Institute of Physics, University of Graz, A-8010 Graz, Austria,\\
\email{ju.jung@uni-graz.at}
}

\maketitle              

\begin{abstract}
We examine the contribution of the pion cloud to the electromagnetic $N \rightarrow \Delta$ transition form factors within a relativistic hybrid constituent-quark model. In this model baryons consist not only of the $3q$ valence component, but contain, in addition, a $3 q \pi$ non-valence component. We start with constituent quarks which are subject to a scalar, isoscalar confining force. This leads to an $SU(6)$ spin-flavor symmetric spectrum with degenerate nucleon and Delta masses. Mass splitting is caused by pions which are assumed to couple directly to the quarks. The point-form of relativistic quantum mechanics is employed to achieve a relativistically invariant description of this system. The $N \rightarrow \Delta$ transition current is then determined from the one-photon exchange contribution to the $\Delta$ electroproduction amplitude. We will give predictions for the ratios $R_{EM}$ and $R_{SM}$ of electric to magnetic and Coulomb to magnetic form factors, which are supposed to be most sensitive to pion-cloud effects.
\end{abstract}
\section{Formalism}
\begin{figure}[tb]
\center{\includegraphics[width=0.30\textwidth]{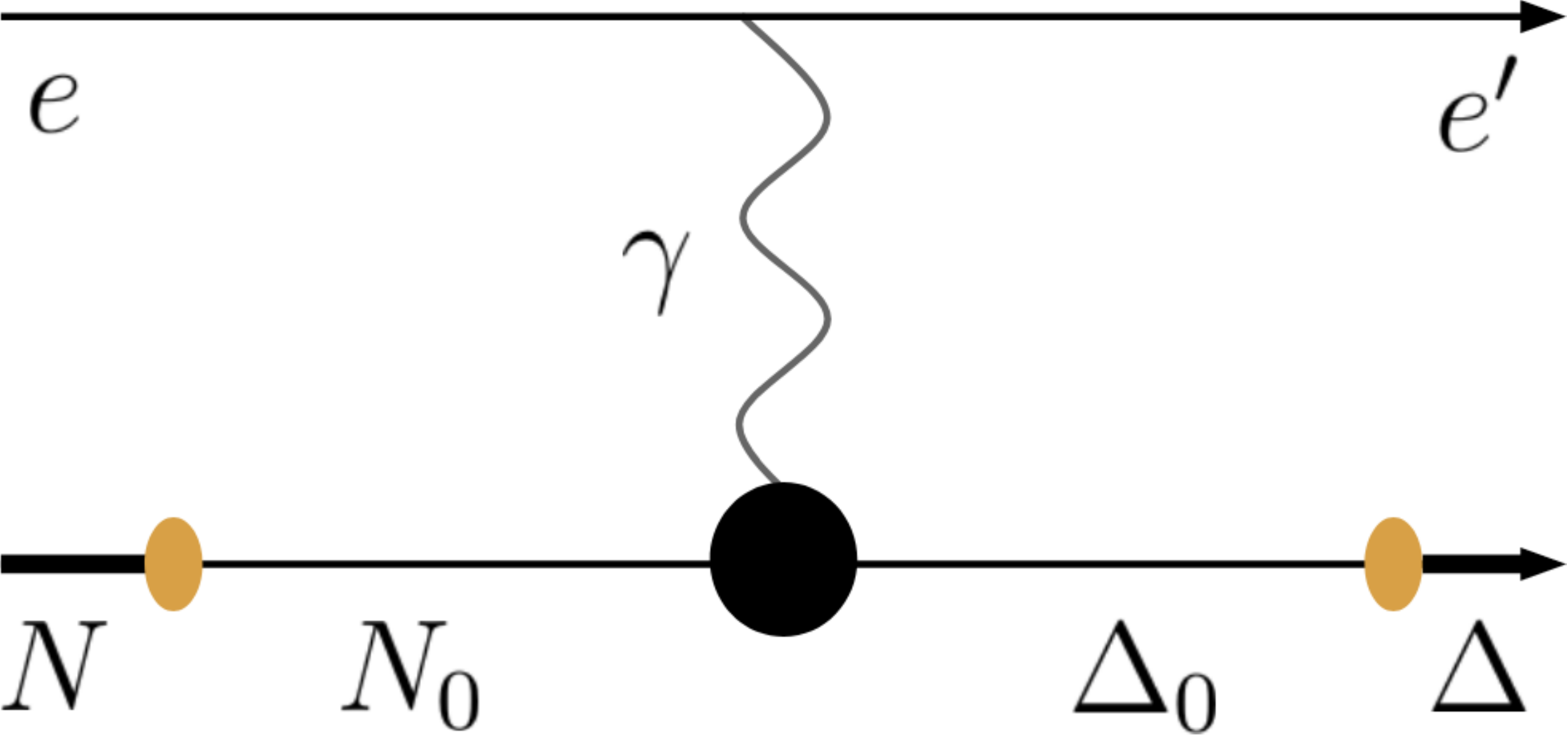}\\ \medskip
\includegraphics[width=0.30\textwidth]{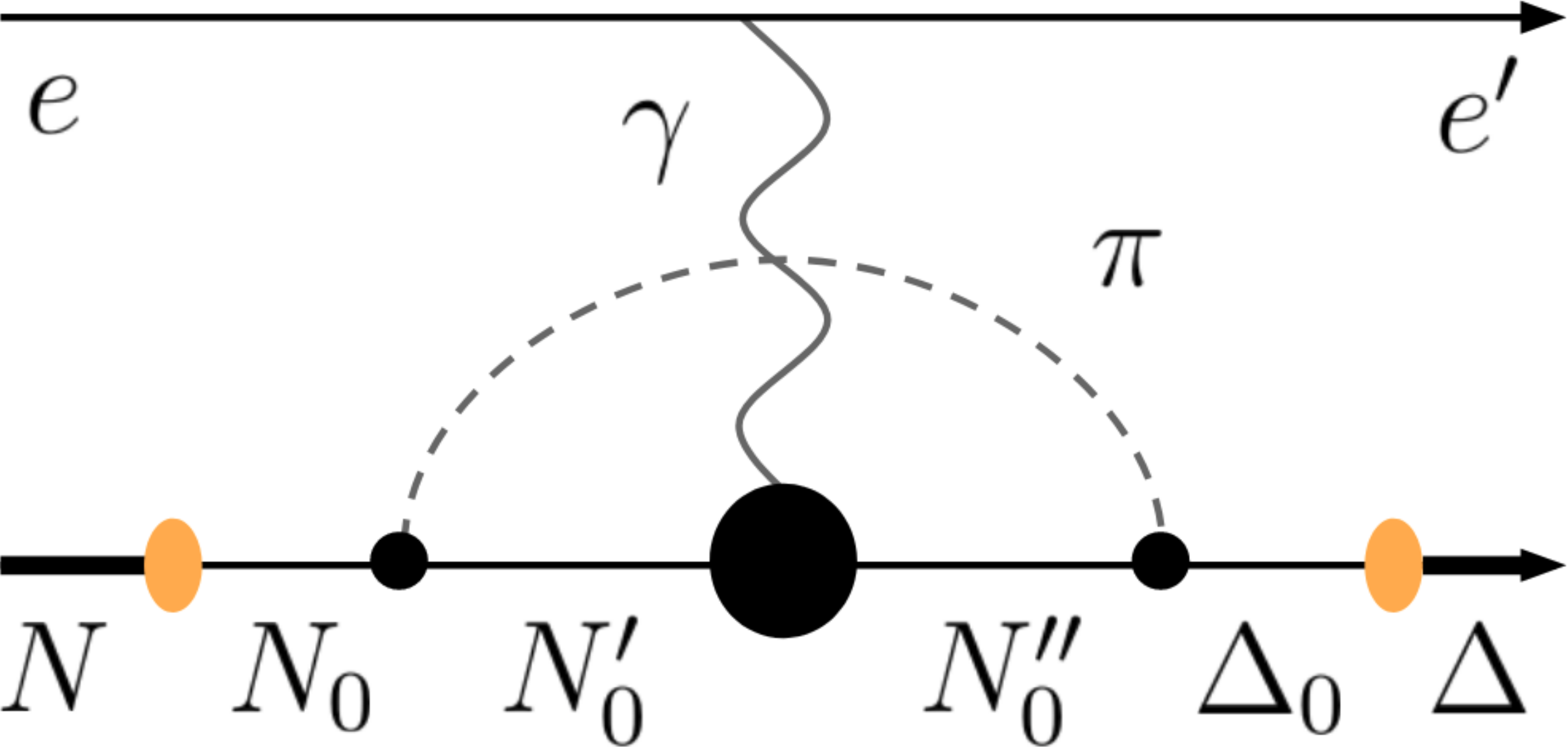} \hspace{2.0cm} \includegraphics[width=0.30\textwidth]{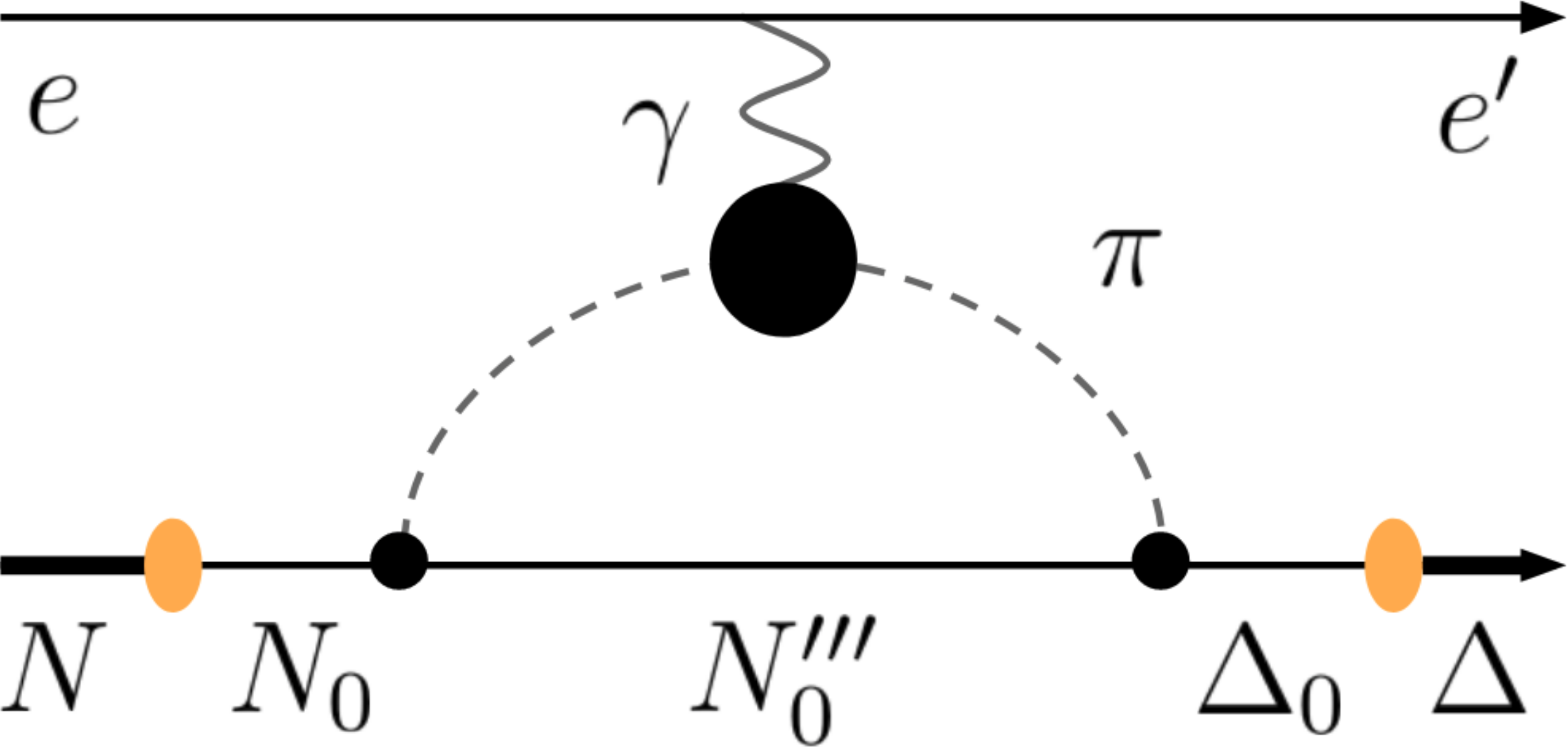}
}
\vspace{-0.3cm}
\caption{The three graphs contributing to electroexcitation of the $\Delta$ resonance in the presence of a pion cloud. The big blobs represent electromagnetic (transition) form factors involving the bare nucleon $N_0$ and the bare Delta $\Delta_0$. The small black blobs represent strong form factors at the $\pi N_0 N_0$, $\pi N_0 \Delta_0$ vertices. All these form factors are determined by the valence-quark wave functions of the bare baryons. A vertex form factor, calculated within a constituent-quark model~\cite{Biernat:2009my} and the same approach as used here, is also assumed at the pion-photon vertex.}
\label{fig:1}
\vspace{-0.5cm}
\end{figure}

For a proper relativistic description of the $N\rightarrow\Delta$ transition form factors
we make use of point-form relativistic quantum mechanics in connection with
the Bakamjian-Thomas construction~\cite{Kli:2018}. Like in previous work~\cite{Biernat:2009my,Biernat:2014dea} we use this framework to determine the one-photon-exchange amplitude for $e^- p\rightarrow e^- \Delta^+$ scattering. From this scattering amplitude we extract the electromagnetic $p\rightarrow \Delta^+$ transition current and determine the form factors by means of a covariant analysis of the transition current.
Thereby both, the nucleon and the Delta are assumed to consist of a $3q$ and a $3q$+$\pi$ component and, in addition to the dynamics of electron and quarks, the dynamics of the photon and the pion are fully taken into account. This is accomplished by means of a multichannel formulation that comprises all states which can occur during the scattering process (i.e. $|3q, e \rangle$, $|3q, \pi, e \rangle$, $|3q, e, \gamma \rangle$, $|3q, \pi, e, \gamma \rangle$). After reducing the mass eigenvalue equation for this system of coupled states to the $3qe$-component, one ends up with an eigenvalue equation of the form
\begin{equation}\label{eq:Mphys}
\left[\hat{M}_{3qe}^{\mathrm{conf}} +\hat{K}_\pi(\sqrt{s}-\hat{M}_{3q\pi e}^{\mathrm{conf}} )^{-1} \hat{K}_\pi^\dag + \hat{V}_{1\gamma}^{\mathrm{opt}}(\sqrt{s})\right] \ket{\psi_{3q e}} = \sqrt{s} \, \ket{\psi_{3q e}} \, ,
\end{equation}
where $\hat{V}_{1\gamma}^{\mathrm{opt}}(\sqrt{s})$ is the 1$\gamma$-exchange optical potential, $\sqrt{s}$ the invariant mass of the scattering system and $\hat{K}_\pi$ the $qq\pi$ vertex operator. We assume an instantaneous scalar and isoscalar confining force between the quarks, which enters $\hat{M}_{3q(\pi)e}^{\mathrm{conf}}$.  The invariant 1$\gamma$-exchange amplitude for electroproduction of the Delta is now obtained by sandwiching $\hat{V}_{1\gamma}^{\mathrm{opt}}(\sqrt{s})$ between (the valence component of) physical electron-nucleon  $\ket{eN}$ and electron-Delta $\ket{e\Delta}$ states , i.e. eigenstates of $[  \hat{M}_{3qe}^{\mathrm{conf}} +\hat{K}_\pi (\sqrt{s}-\hat{M}_{3q\pi e}^{\mathrm{conf}} )^{-1} \hat{K}_\pi^\dag ]$. The crucial point is now to observe that, due to instantaneous confinement, propagating intermediate states do not contain free quarks, they rather contain bare nucleons $N_0$ or bare Deltas $\Delta_0$ (or corresponding excitations, which are neglected in our calculations). The bare particles are eigenstates of the pure confinement problem. This allows us to rewrite the scattering amplitude in terms of pure hadronic degrees of freedom with the quark substructure being hidden in strong and electromagnetic vertex form factors of the bare baryons. This is graphically represented in Fig.~\ref{fig:1}.

For scalar, isoscalar confinement the masses of the bare nucleon and Delta are the same, $m_{N_0}=m_{\Delta_0}=:m_0$, and also the three-quark wave functions coincide due to $SU(6)$ spin-flavor symmetry. Instead of choosing a particular confining interaction we therefore rather parameterize the three-quark wave function of $N_0$ and $\Delta_0$ by means of a Gaussian.
Knowing the bare mass $m_0$, the (pseudovector) pion-quark coupling $f_{\pi q q}$ and the constituent-quark masses $m_u=m_d=:m_q$, one can first calculate the strong couplings and form factors at the $\pi N_0 N_0$, $\pi N_0 \Delta_0$ and $\pi N_0 \Delta_0$ vertices and in the sequel the renormalization effect of pion loops on the nucleon and Delta mass. Fixing the constituent-quark mass $m_q$ in advance, we have varied the remaining three parameters ($m_0$, $f_{qq\pi}$ and $\alpha$) by means of a self-consistent procedure such that the solution of a mass-eigenvalue problem analogous to Eq.~(\ref{eq:Mphys}) (just without electron and photon) gives the physical nucleon and Delta masses. Our resulting parameters for two common choices of the constituent-quark mass are given in Tab.~\ref{tab:param}. A more detailed account of the parameter fixing can be found in Ref.~\cite{Jung:2017cpy}.

\begin{table}[t!]\label{tab:param}
\begin{center}
\begin{tabular}{|l | cccc|}
\hline
&$m_q$ [GeV]&$f_{qq\pi}$&$m_0$ [GeV]&$\alpha$\\
\hline
Model I&0.263&0.8067&1.380&2.660\\
Model II&0.340&0.7565&1.390&2.585\\
\hline
\end{tabular}
\end{center}
\caption{Model parameters for two common choices of the constituent-quark mass $m_q$.}
\vspace{-0.8cm}\end{table}

What is still necessary to calculate the leading-order electroproduction amplitude, as depicted in Fig.~\ref{fig:1}, are the electromagnetic (transition) form factors of the bare baryons. These are obtained from the first graph in Fig.~\ref{fig:1} by identifying the bare and the physical baryons.
As one would expect, the one-photon exchange amplitude for $e B_0\rightarrow e B_0^\prime$ scattering can be written as (covariant) photon  propagator times electron current contracted with the baryonic current, $\mathcal{M}_{1\gamma}^{e B_0\rightarrow e B_0^\prime}\propto j_{e\mu} I^\mu_{B_0\rightarrow B_0^\prime}/Q^2$. This allows to extract a microscopic expression for the baryonic current $I^\mu_{B_0\rightarrow B_0^\prime}$. The form factors are then obtained by means of a general covariant decomposition of $I^\mu_{B_0\rightarrow B_0^\prime}$. The resulting model current $I^\mu_{B_0\rightarrow B_0^\prime}$, however,can be afflicted by unphysical contributions (depending on the electron momentum) which are partly eliminated by extracting the form factors in the infinite momentum frame. But problems with the \lq\lq angular condition\rq\rq\ may still persist. This deficiency can be traced back to problems with cluster separability inherent in the Bakajian-Thomas construction~\cite{Kli:2018}. A more detailed account of how we deal with these problems in case of the $N\rightarrow\Delta$ transition current can be found in Ref.~\cite{Jung:2018lmk}.

\vspace{-0.3cm}
\section{Results and Discussion}
\begin{figure}[t!]\label{fig1}
\includegraphics[width=0.5\textwidth]{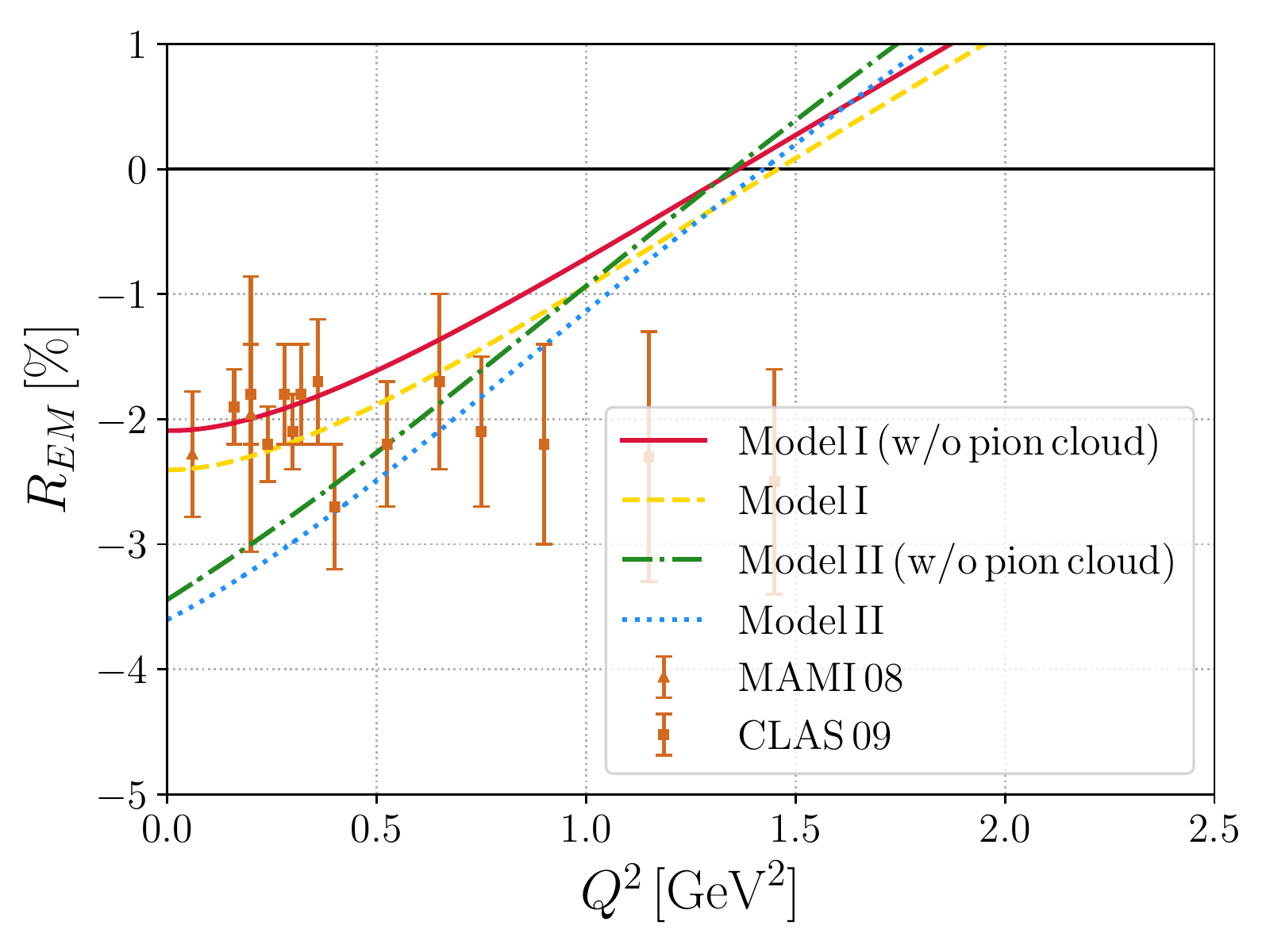}\hfill\includegraphics[width=0.5\textwidth]{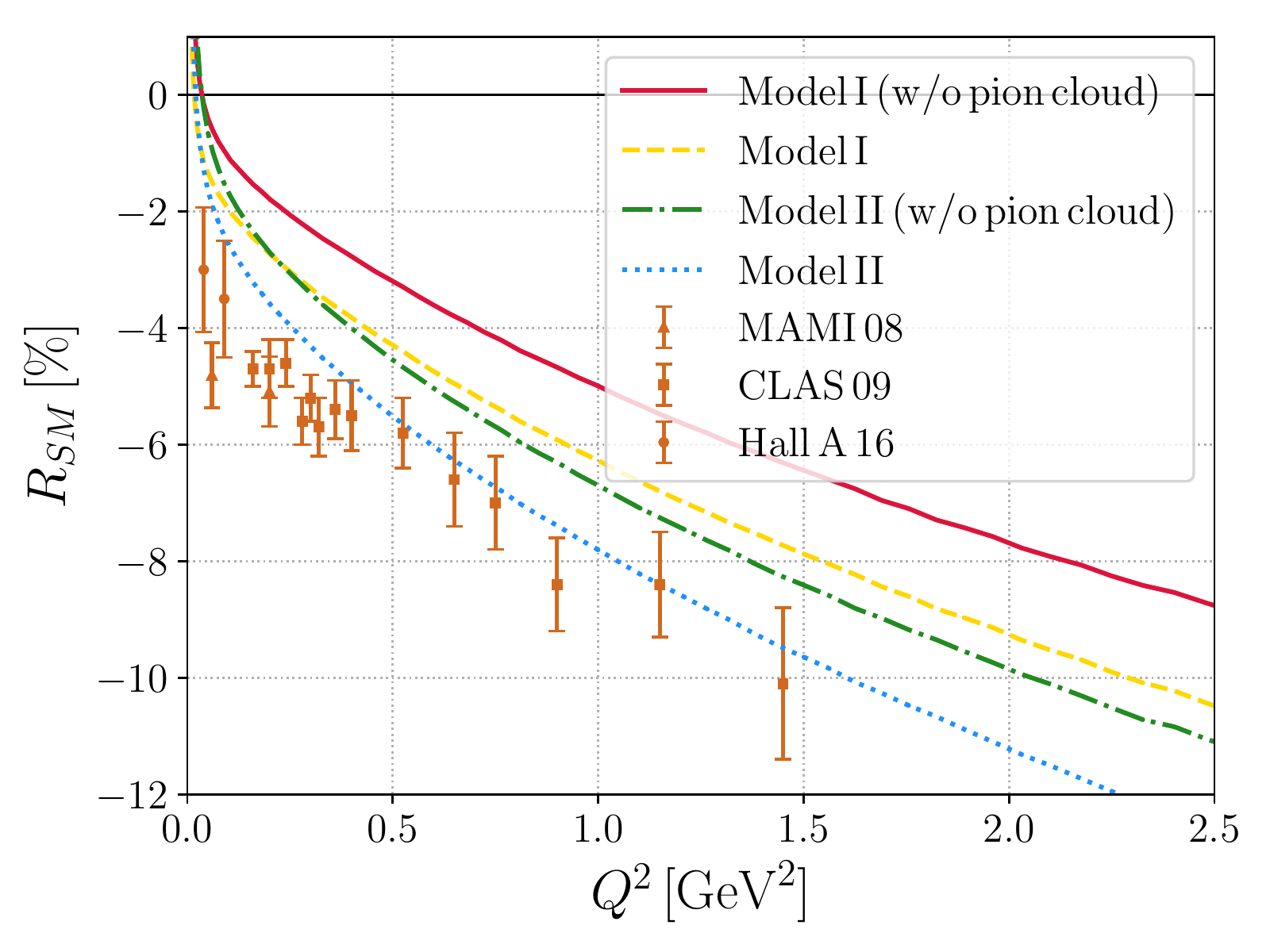}
\vspace{-0.7cm}\caption{Our model predictions for the ratios $R_{EM}$ and $R_{SM}$ (for the definition, see text) as compared with experimental data from MAMI~\cite{Sparveris:2008jx}, CLAS~\cite{Aznauryan:2009mx} and JLab~\cite{Blomberg:2015zma}. Model I and model II refer to the two parameterizations given in Tab.~\ref{tab:param}. Curves labelled \lq\lq w/o pion cloud\rq\rq\ refer to the conventional constituent-quark model without pion cloud, whereas the other results are those for the full calculation, including the pion cloud. }\vspace{-0.2cm}
\end{figure}

With the strong and electromagnetic form factors of the bare baryons we are now able to calculate the pion-loop contributions to the electromagnetic $p\rightarrow \Delta^+$ transition form factors. We account only for the $N_0 \pi$ component in the physical nucleon and Delta, but neglect the $\Delta_0\pi$ component. There is some evidence from phenomenological hadronic models that an $SU(6)$ spin-flavor symmetric model like ours would overestimates the $\Delta_0\pi$ component considerably. A common choice for electromagnetic $p\rightarrow \Delta^+$ transition form factors is the one suggested by Jones and Scadron~\cite{Jones:1972ky}. Pion-cloud effects are most visible in the small form factors $G_E^\ast$ and $G_C^\ast$. What is often plotted are the ratios $R_{EM}=-G_E^\ast/G_M^\ast$ and $R_{SM}=-(Q_+ Q_-/(4 m_\Delta^2))\,G_C^\ast/G_M^\ast$, where $Q_{\pm}=\sqrt{(m_\Delta\pm m_N)^2+Q^2}$. These are shown in Fig.~\ref{fig1} for the two parameterizations of our model given in Tab.~\ref{tab:param}. Our results compare with the outcome of other theoretical predictions coming from constituent-quark models~\cite{Cardarelli:1995dc,Ramalho:2008dp,Sanchis-Alepuz:2017mir}.
For $Q^2\gtrsim 0.5$~GeV$^2$ our predictions for $G_{M}^{*}$ agree well with the data, for vanishing $Q^2$, however, we underestimate the data by about $15\%$ for model I and about $25\%$ for model II. Model I works also better for $R_{EM}$, whereas a better reproduction of $R_{SM}$ is achieved with model II. The pion-cloud contribution is clearly visible in both ratios and it goes into the right direction.

There is, of course, room left for improvement. One should keep in mind that our starting point was $SU(6)$ spin-flavor symmetry for the bare baryons. One could, e.g., think of introducing $SU(6)$ symmetry-breaking effects right from the beginning, which lead to different masses and wave functions for the bare nucleon and Delta. This perhaps will also lead to a more reasonable probability for finding the $\pi \Delta_0$ component in the physical nucleon and Delta. Contributions  from $\pi \Delta_0$ intermediate states could then also help to improve agreement with data. It is the topic of future work to find out, whether $SU(6)$-symmetry breaking effects on the bare baryon level (in addition to pion-cloud effects) suffice to improve the agreement with data, or whether, e.g., an explicit $d$-wave contribution to the $\Delta$ wave function, as it is asserted by several authors (see, e.g., \cite{Ramalho:2008dp}), will be necessary to achieve a satisfactory reproduction of data.

\medskip

{\bf Acknowledgement:}
J.-H. Jung acknowledges the support of the Fonds zur F\"orderung der wissenschaftlichen Forschung in \"Osterreich (Grant No. FWF DK W1203-N16).\vspace{-0.3cm}

%
%

\end{document}